\documentclass[twocolumn,showpacs,preprintnumbers,amsmath,amssymb]{revtex4-1}
\pdfoutput=1
\usepackage{graphicx}
\usepackage{bm}
\usepackage{times}
\usepackage{slashed}
\usepackage{color}
\usepackage{slashed}
\usepackage{amsmath}
\usepackage{amsthm}
\usepackage{subfigure}

\newcommand \gmu { $g-2_{\mu}$ }
\def\br{\begin{eqnarray}}
\def\er{\end{eqnarray}}
\def\be{\begin{equation}}
\def\ee{\end{equation}}

\def\({\left(}
\def\){\right)}
\def\<{\left\langle}
\def\>{\right\rangle}

\begin{document}
%
%
\title{ {\color{blue} Composite Higgs Models, Technicolor and The Muon Anomalous Magnetic Moment}}

\author{A. Doff$^{a}$, Clarissa Siqueira$^{b}$}
\affiliation{$^a$ Universidade Tecnol\'ogica Federal do Paran\'a - UTFPR - DAFIS,  Av Monteiro Lobato Km 04,84016-210, Ponta Grossa, PR, Brazil\\$^b$ Departamento de Fisica, Universidade Federal da Paraiba,
Caixa Postal 5008, 58051-970, Joao Pessoa - PB, Brazil.}

\date{\today}
\begin{abstract} 
We revisit the muon magnetic moment (g-2) in the context of Composite Higgs models and Technicolor, and provide general analytical expressions for computing the muon magnetic moment stemming from new fields such as, neutral gauge bosons, charged gauge bosons, neutral scalar, charged scalars, and exotic charged leptons type of particles. Under general assumptions we assess which particle content could address the \gmu excess. Moreover, we take a conservative approach and derive stringent limits on the particle masses in case the anomaly is otherwise resolved and comment on electroweak and collider bounds. Lastly, for concreteness we apply our results to a particular Technicolor model.

\end{abstract}
\maketitle
\section{Introduction}

The Standard Model (SM) of elementary particles is in excellent agreement with the experimental data and has been able to endure electroweak precision data throughout the years.  The nature of the electroweak symmetry breaking is one of the most important problems in particle physics, and the $125$ GeV new resonance discovered at the LHC \cite{LHC} has many of the characteristics expected for the Standard Model (SM) Higgs boson.
Despite its success, there are observational reasons to believe that the standard model is not the whole story, such as dark matter, neutrino masses, and more fundamental ones such as the hierarchy problem. 

Here we try to asses some models concerning the muon anomalous magnetic moment that are capable of addressing the aforementioned matters.

The muon magnetic muon is one of the most precisely measured observables in particle physics. There is an old discrepancy of $3.6 \sigma$ between the theoretical SM contribution to g-2 and its measured value  \cite{PDG}. This deviation gave rise to numerous new physics effects speculated to be plausible answers to the exciting excess. One of the striking features of the muon magnetic moment is its sensitivity to new physics effects coming from low to very high energy scale models. Moreover, it is fair to say that the majority of the extensions to the SM give sizeable corrections to g-2 in a certain region of parameter space. Albeit, due to the embedded theoretical and experimental uncertainties surrounding this quantity a conservative approach is needed in order to derive robust results.

Currently, the difference, $a^{exp}_{\mu}-a^{SM}_{\mu}= (296 \pm 81) \times 10^{-11}$, which corresponds to $3.6\sigma$ \cite{fermilabproposal}, can be reduced up to $2.4\sigma$ if one used $\tau$ data in the hadronic corrections \cite{PDG}. Thus, it is clear that a large improvement in the theoretical calculation from SM is demanded before claiming a new physics discovery.

Therefore, we will take a conservative approach in this work. We will discuss the possibility of addressing g-2 with new fields as well as derive bounds on the particles masses by enforcing their contributions to be within the error bars reported by the experiments, having in mind interactions that appear in Composite Higgs models (CHM) and Technicolor models (TC) for the following reasons:(i)  CHM  provide a plausible solution to the hierarchy problem since the Higgs sector is replaced by a new and strongly coupled sector. The strong sector contains a global symmetry, which is then spontaneously broken at a scale $\Lambda$ and the Higgs is identified as one of the  Nambu-Goldstone bosons \cite{Barnard:2014tla}. CHM can also have explicit global symmetry breaking by linear couplings of SM field to operators in the strong sector, thus inducing an electroweak symmetry breaking and generating the Higgs mass \cite{Barnard:2014tla}. (ii) Alternatively, inspired in QCD, Technicolor was a theory invented to provide a natural and consistent quantum-field theoretic description of electroweak (EW) symmetry breaking,  without elementary scalar fields. TC are based on the introduction of a new strong interactions, where in these theories the Higgs boson is a composite field of the so called technifermions. The beauty of TC as well as its problems are clearly summarized in Refs.\cite{Sannino:2009za,Lane:2000pa,Hill:2002ap}. In particular, the model described in Ref.\cite{Doff:2010pp} the electroweak symmetry is broken dynamically by a technifermion condensate generated by the $SU(2)_{TC}$ Technicolor (TC) gauge group \cite{Doff:2013wba}.

One of the most distinct differences between those models is that in CHM the electroweak symmetry is not directly broken due to a fermion condensate.  Albeit, the fermion condensates become strong at a high scale, say $\Lambda = 10 TeV$, breaking a global symmetry that results into a heavy pseudo-scalar that mixes with the Higgs that needs to be introduced to generate the Yukawa lagrangians and generate fermion masses. In other words, the electroweak symmetry in such models is simply the vacuum alignment produced by the Higgs and Pseudo-scalar mixing. This solves the the hierarchy problem because at a scale larger than 10 TeV there is a condensate,  whereas, in Technicolor  the condensate of technifermions that have the quantum number $SU(2) \times U(1)$ condense the give masses to the gauge bosons. Despite those subtle differences in the pattern of symmetry breaking, Techinicolor and CHM share similarities as far as the muon magnetic moment is concerned.

In summary, motivated by interesting aspects of the both CHM and TC models and the g-2 discrepancy we will revisit the g-2 in terms of simplified lagrangians which rise in those models, and lastly apply our findings to a particular technicolor model. We begin by discussing the muon magnetic moment.

\section{Muon Magnetic Moment}

\begin{figure*}[!t]
\centering
\subfigure[\label{feymann4}]{\includegraphics[scale=0.5]{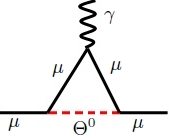}}
\subfigure[\label{feymann5}]{\includegraphics[scale=0.5]{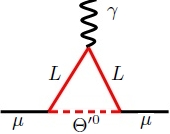}}
\subfigure[\label{feymann1}]{\includegraphics[scale=0.5]{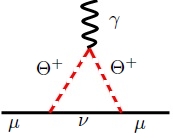}}
\subfigure[\label{feymann7}]{\includegraphics[scale=0.5]{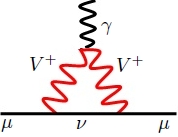}}
\subfigure[\label{feymann8}]{\includegraphics[scale=0.5]{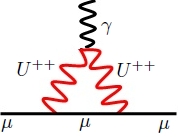}}
\subfigure[\label{feymann9}]{\includegraphics[scale=0.5]{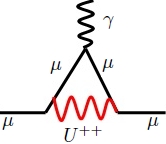}}
\caption{Feynmann diagrams that contribute to the muon magnetic moment.}
\end{figure*}

As important as charge, mass, spin and lifetime of a given particle, the magnetic moments are fundamental quantities. On the classical level, an orbiting particle with electric charge e and mass m exhibits a magnetic dipole moment given by $\overrightarrow{\mu}= e/2m \overrightarrow{L}$, where $\overrightarrow{L}$ is the orbital angular momentum. To measure the magnetic moment we need the presence of a magnetic field (B) because the observable Hamiltonian goes as -$\overrightarrow{\mu}$. $\overrightarrow{B}$. However, this classical view is abandoned for particles and the magnetic moment, which is intrinsic, is obtained replacing the angular momentum by the spin, in such way that now $\overrightarrow{\mu} = -g\, e/(2m) \overrightarrow{S}$. The deviation from the Dirac value g/2 = 1, obtained on the classical level, is $a_{\mu}= (g-2)/2$, so called muon anomalous magnetic moment. The experimental and theoretical results are reported in terms of it. 
The SM contributions to $a_{\mu}$ are divided into three parts: electromagnetic (QED), electroweak (EW) and hadronic ones. The QED part consist of all photonic and leptonic contributions and has been evaluated up to 4-loops, whereas the EW involves the SM bosons ($W^{\pm},Z$,H) and has been computed up to three loops \cite{PDG}. Lastly, the hadronic contributions has to do with quarks running in the loops. Due to the large fermion masses involved, the hadronic contributions carry the largest uncertainties, in particular the hadronic vacuum polarization, which is calculated from $e^+e^- \rightarrow {\rm hadrons}$, or $\tau \rightarrow {\rm hadrons}$ data \cite{PDG}, and the hadronic light-by-light scattering, which currently cannot be determined from  data \cite{lightlight}, and give rise to the most relevant errors. In summary, the SM expected contribution to the muon anomalous magnetic moment is $a_{\mu}^{SM} = (116591785 \pm 51) \times 10^{-11}$ \cite{fermilabproposal}.
Although, the E821 experiment at Brookhaven National Laboratory, which studied the precession of muon and anti-muon in a constant external magnetic field as they circulated in a confining storage ring, reported the following value $a_{\mu}^{E821}= (116592080 \pm 63) \times 10^{-11}$ \cite{Bennett:2004pv,Bennett:2006fi}. Thus, 

\begin{equation}
\Delta a_{\mu} (E821 -SM) = (295 \pm 81) \times 10^{-11},
\label{deltaa}
\end{equation}which results into a $3.6\sigma$ excess. 

Out of $\pm 81$ error, $\pm 51 \times 10^{-11}$ is associated to theoretical uncertainties. In particular, $\pm 39 \times 10^{-11}$ stems from the lowest-order hadronic contribution and $\pm 26 \times 10^{-11}$ rises from hadronic light-by-light contributions \cite{fermilabproposal}. An important effort has been put forth trying to reduce the theoretical and experimental errors. In the light of the g-2 experiment at FERMILAB both uncertainties are expected to be substantially reduced and bring the total error down to  $\pm 34 \times 10^{-11}$ \cite{fermilabproposal,g2muontheory2}. In our figures, we exhibit a dark (light) gray band that delimits the mass range which accommodates the g-2 anomaly, and two red horizontal lines, where the solid (dashed) refers to the current (projected) $1\sigma$ bound based on the present (expected) $\pm 81$ ($\pm 34$)  error bar.

The muon magnetic moment is  tightly related to the flavor violating $\mu \rightarrow e \gamma$ decay. Thus our limits are also strongly correlated to those rising from flavor violating decays. Current data imposes ${\rm BR}( \mu \rightarrow e \gamma) < 5.7 \times 10^{-13}$ \cite{Adam:2013mnn}, with,

\begin{equation}
{\rm BR}( \mu \rightarrow e \gamma) \simeq 6. 34 \times 10^{-7} \left(\frac{1 TeV}{\Lambda}\right)^4 \lambda_{\mu e}^2,
\end{equation}where $\Lambda$ refers to the new physics scale and $\lambda_{\mu e}$ to the flavor violating coupling constant \cite{Cirigliano:2005ck}. Notice that for new physics processes which occur at the TeV scale rather dwindled couplings are required. Although, one can in principle postulate the presence of new symmetries or simply make use of suppressed non-diagonal couplings \cite{Vicente:2014wga,Restrepo:2013aga,Frigerio:2014ifa,
Chakrabortty:2012vp,Rodejohann:2011vc,Dev:2015uca}. Hereafter, we focus on the muon magnetic moment only, but the reader should keep in mind that competitive $\mu \rightarrow e \gamma$ bounds might exist. Now we have reviewed the status of the muon anomalous magnetic moment, we further discuss general features of Composite Higgs(CH) and Technicolor models.  

\section{Composite Higgs and Technicolor Models}

\begin{figure}[!t]
\centering
\includegraphics[scale=0.55]{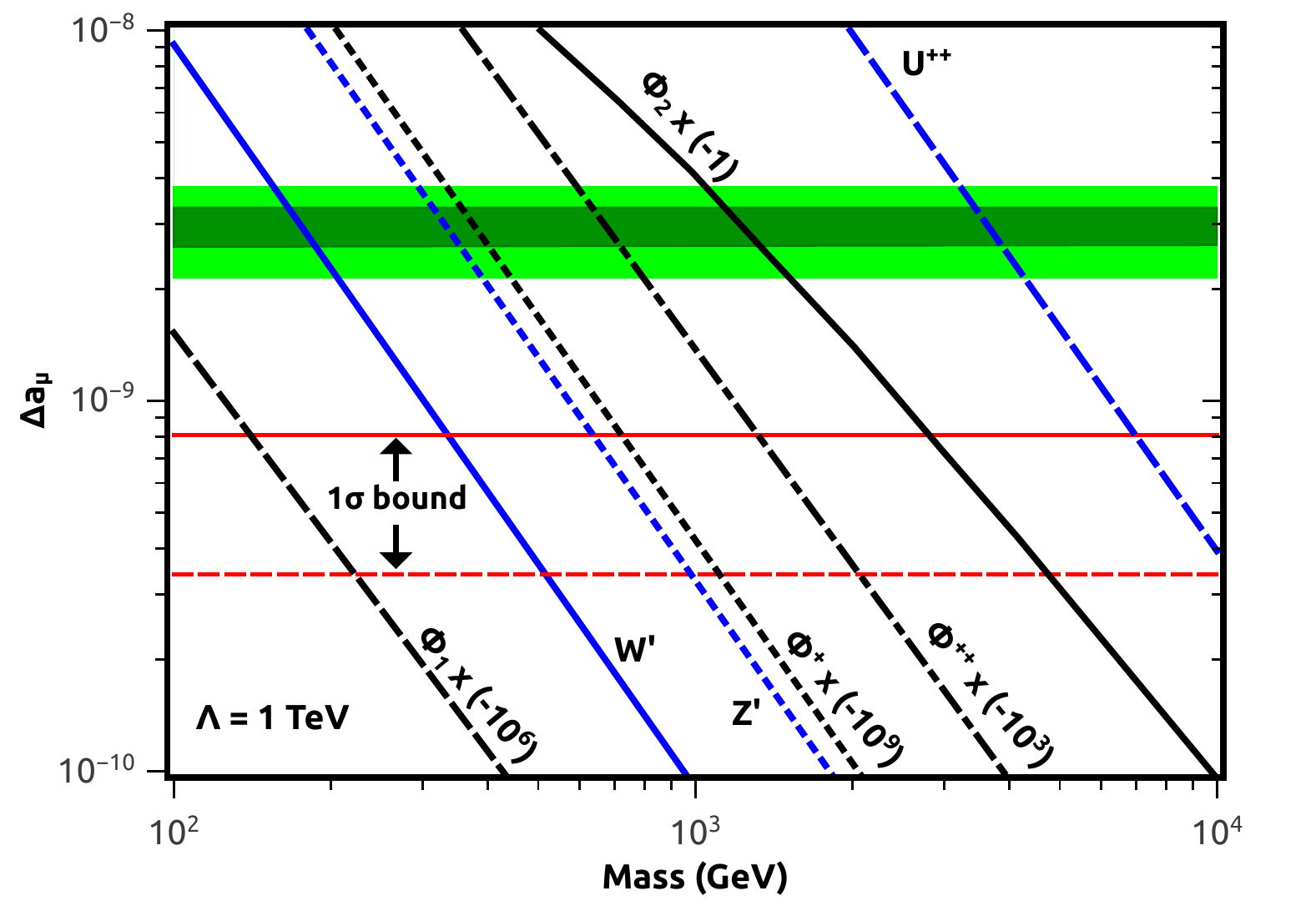}
\caption{Individual corrections to the muon magnetic moment as function of the boson masses for $\Lambda = 1$~TeV and $m_L=100$~GeV. The green band delimits the current (light) and projected (dark) sensitivity of the muon magnetic moment.The solid and dashed red lines represent the current and project $1\sigma$ limit in case the anomaly is otherwise resolved.}
\label{plotinde}
\end{figure}

\begin{figure}[!t]
\centering
\includegraphics[scale=0.55]{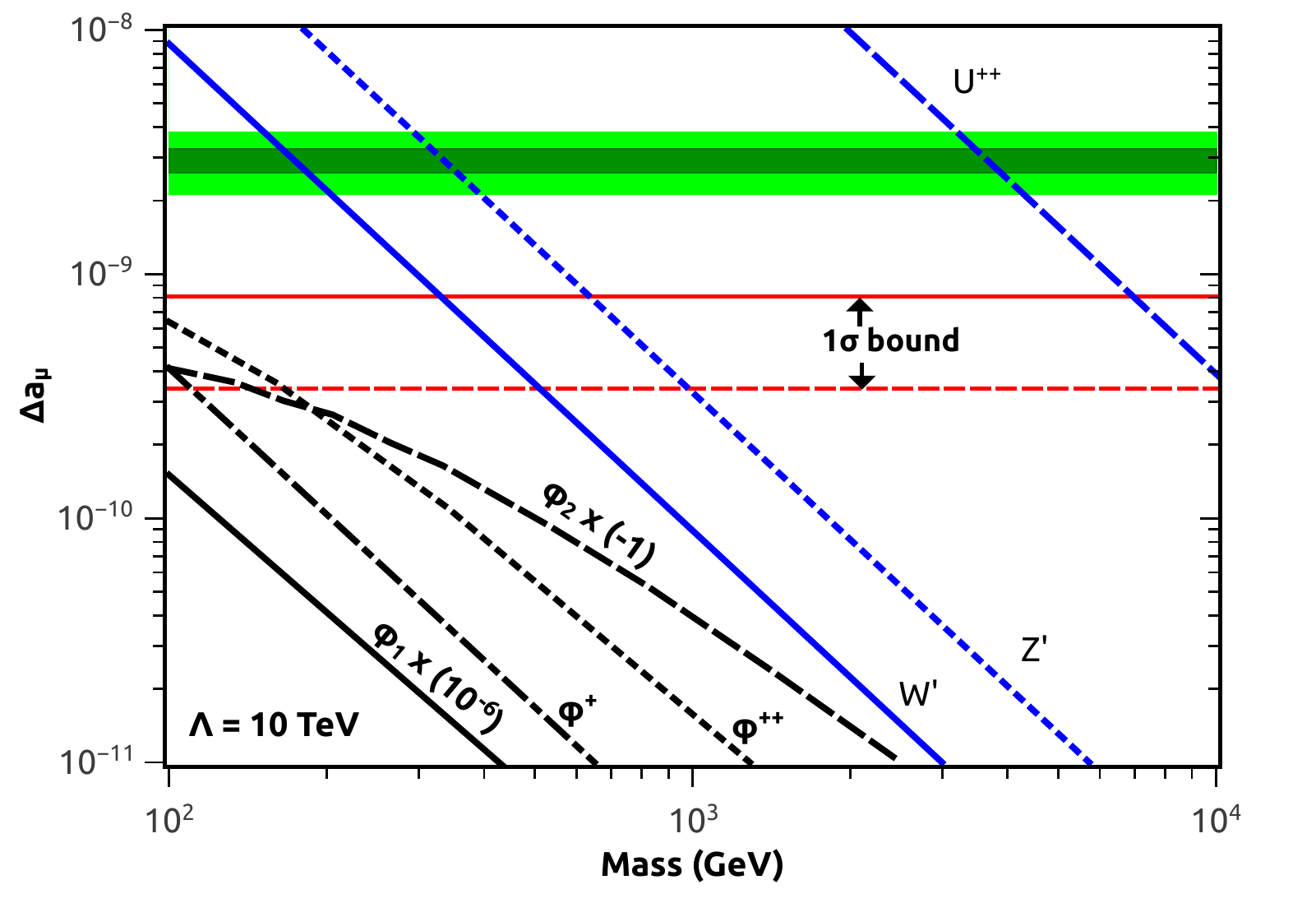}
\caption{Individual corrections to the muon magnetic moment as function of the boson masses for $\Lambda = 10$ TeV and $m_L=100$ GeV.}
\label{plotinde2}
\end{figure}

Composite Higgs models (CHM) are extensions of the SM where the Higgs boson is a bound state of new strong interactions. These models are arguably the leading alternative to supersymmetric models, since they provide an explanation to the hierarchy problem. One of the main features of CHM are the existence of new particles with masses at the TeV scale that are excitations of the composite Higgs. Those particles can be potentially produced and discovered in the foreseeable future at the LHC. Moreover, such particle could produce deviations from the SM predictions in low energy observables such as the muon magnetic moment, which is the focus of this work. There are various ways to generate the Higgs boson, but CHM can be broadly divided in two categories: (i) Higgs is a generic  composite bound state of strong dynamics (TC); (ii) Higgs is a Goldstone boson of spontaneous symmetry breaking(CHM). (See \cite{Hagedorn:2011pw,Marzocca:2012zn,Caracciolo:2012je,
Fonseca:2015gva,Carena:2014ria,vonGersdorff:2015fta} for recent phenomenological works on CHM)

The possibility that the Higgs boson is a composite state instead of an elementary one is more akin to the phenomenon  of spontaneous symmetry breaking that originated from the effective Ginzburg-Landau Lagrangian, which can be derived from the microscopic BCS theory of superconductivity describing the electron-hole interaction (or the composite state in our case). This dynamical origin of the spontaneous symmetry breaking has been discussed with the use of many models, with technicolor being the most popular one.  The early technicolor models suffered from problems such as flavor changing neutral currents (FCNC)  and contribution to electroweak observables in disagreement with experimental data, as can be seen in the reviews of Ref.\cite{Hill:2002ap,Sannino:2009za}. Nevertheless, the TC dynamics may be quite different from the known strong interaction theory, i.e. QCD.  This fact has led to the walking TC proposal \cite{Catterall:2007yx}, where the incompatibility with the experimental data has been resolved, making the new strong interaction almost conformal and changing appreciably its dynamical behavior. In the latter TC theory the technifermion self-energy will acquire large current masses, and subsequently the pseudos-Goldstone bosons formed with these ones. An almost conformal TC theory can be obtained when the fermions are in the fundamental representation, introducing a large number of TC fermions $(nF)$, leading to an almost zero $\beta$ function and flat asymptotic coupling constant. This procedure may induce a large S parameter incompatible with current electroweak measurements though. The perturbative expression to the S parameter (in the massless limit) is given by  
\be 
S = \frac{1}{6\pi}N_d d(r)
\ee 
\noindent where $N_d$ is de number of left-handed electroweak technidoublets, and $d(r)$  is the dimension of the technifermion representation. Data require the value of the S parameter to be less than about 0.3, TC models with fermions in other representations than the fundamental one, such as Minimal\cite{Foadi:2007ue} (MWT) and Ultraminimal \cite{Ryttov:2008xe} (UMT) TC models are viable models that accommodate the measured S parameter. As for flavor changing neutral-current (FCNC) processes, the vector bosons that mediate generation-changing transitions must have large masses $\sim 10^3$~TeV. Moreover, corrections from heavy fermions (top) and pseudo-scalars which are set by the Techinicolor symmetry breaking scale, require the latter to be in the ballpark of $\sim 1$~TeV. Additional limits arise if one wants to incorporate dark matter particles \cite{Fonseca:2015gva}. Studies using high dimensional operators have been performed and shown that with a $\sim$ TeV symmetry breaking scale such model might reproduce the correct relic abundance while avoiding direct \cite{Akerib:2013tjd} and indirect detection  bounds \cite{Hooper:2012sr}. See for a review concerning current LHC data \cite{Csaki:2015hcd}.

Setting aside those subtleties, CH and TC models share similar features as far as the muon magnetic moment is concerned. Some models postulate the existence of, not limited to, neutral vectors, charged vectors, neutral pseudo-scalars $\phi^0$, charged scalars $\phi^+$, exotic charged leptons (L), and even doubly charged gauge bosons (see Ref. \cite{Beneke:2012ie,Redi:2013pga,Antipin:2014mda}). We point out that precision-electroweak observables such as the oblique parameters S and T result into a robust bound on the scale of symmetry breaking of (CH) namely,  $\Lambda > 0.8-5.5$ TeV \cite{Barnard:2014tla}. The precise limit strongly depends on the particular details of the model \cite{Barnard:2014tla}. In the context of TC models the precision-electroweak parameters and constraints from (FCNCs) processes restrict TC models to a specifc dynamic, walking TC models \cite{Sannino:2009za} in our case,  the contribution due to the  TC sector should still lead to a value to the S parameter compatible with the experimental data \cite{Csaki:2015hcd}. That being said, here we derive analytical expressions to compute the muon magnetic moment for several particles that are present in some Technicolor and CHM models in a general setting, assuming then that the possible contributions of these theories  are due to  TeV energy scale. 

 \section{Composite Higgs Models and Technicolor Contributions to  Muon Anomalous Magnetic Moment}

In general after the chiral symmetry breaking of the  strongly interacting sector a large number of Goldstone bosons can be formed, and only few of these degrees of freedom are absorbed by the weak interaction gauge bosons, which is the case of TC models.  The others may acquire small masses resulting in light pseudo-Goldstone bosons that have not been observed experimentally. However, in the TC models considered in this work  these bosons obtain masses that are large enough to have escaped detection at the present accelerator energies, the possible pseudo-scalars bosons can be listed  according to their different quantum numbers. Some works have devoted attention to the muon magnetic moment in the context of composite higgs models such as \cite{Falkowski:2013jya,Arbuzov:2001yt,Antipin:2014mda,Arbuzov:2001yt}, here we extend those by including a more accurate and general calculation to g-2 stemming from new fields.

\begin{itemize}

\item {\it Pseudo-Scalars}:

As the comment made in the previous section pseudo-scalars   give rise to corrections to muon magnetic moment through the Effective Lagrangian ,

\begin{equation}
\mathcal{L} \supset  \frac{m_{\mu}}{\Lambda} \bar{\mu} \gamma_5 \mu\, \phi_1,
\label{pseudoscalar}
\end{equation}

The correction to g-2 is found to be,

\begin{eqnarray}
&&
\Delta a_{\mu} = \frac{1}{8\pi^2}\frac{m_\mu^2}{ M_{\phi}^2 } \int_0^1 dx \frac{  (m_{\mu} / \Lambda)^2 \ (-x^3) }{(1-x)(1-\lambda^2 x) +\lambda^2 x}\nonumber\\
\label{scalarmuon1}
\end{eqnarray}where $\lambda=m_{\mu}/M_{\phi_1}$ which gives us,
\begin{eqnarray}
&&
\Delta a_{\mu} = \frac{1}{4\pi^2}\frac{m_\mu^2}{ M_{\phi_1}^2 } (m_{\mu} / \Lambda)^2\left[ - \ln \left(\frac{ M_{\phi_1} }{m_{\mu}} \right) +\frac{11}{12}\right] \nonumber\\
\label{scalarmuon3}
\end{eqnarray}in agreement with \cite{Kelso:2013zfa,Queiroz:2014zfa,Freitas:2014pua,Kelso:2014qka}.

In Fig.\ref{feymann4} we exhibit the Feynman diagram for this process. Notice that the additional muon mass suppression is typical in neutral scalars correction to g-2. Hence, are typically neglected. Additionally, we have included the energy scale $\Lambda$ that reflects the Technicolor or CHM symmetry breaking scale. Those two factors suppress the overall correction. Moreover, note that the contribution rising from a pseudo-neutral scalar is always negative and therefore it cannot accommodate the muon magnetic moment excess. We point out that this result is general and applicable to any extension of the SM model. However, we point out that those pseudo-scalars are quite common in CHM and Technicolor models. As aforesaid, the muon mass and the symmetry breaking scale suppressions dwindle general contributions to g-2 stemming from pseudo-scalars. In Fig.\ref{plotinde} the black dashed line is our numerical result for this pseudo-scalar, which has been multiplied by an overall factor of $10^6$ so that we could show it in the figure.

\item {\it Pseudo-Scalars + Charged Lepton}:

Exotic charged Lepton have also been evoked in the models in question \cite{Redi:2013pga,Doff:2010pp,Doff:2012uq,Doff:2007yu} and contribute to g-2 through Fig.\ref{feymann5}. A simplified Lagrangian for this field can be written as,

\begin{equation}
\mathcal{L}  \supset \frac{m_L}{\Lambda} \bar{L} \gamma_5 \mu\, \phi_2
\end{equation}which give rises to,

\begin{eqnarray}
&&
\Delta a_{\mu} = \frac{1}{8\pi^2}\frac{m_\mu^2}{ M_{\phi_2}^2 } \int_0^1 dx \frac{ (m_L / \Lambda)^2 x^2 (1-\epsilon -x) }{(1-x)(1-\lambda^{2} x) +\epsilon^2 \lambda^2 x}\nonumber\\
\label{leptonmuon7}
\end{eqnarray}where $\epsilon = M_{E}/m_{\mu}$ and $\lambda= m_{\mu}/M_{h}$. In the limit $M_{\phi_2} \gg M_L$ we get,
\begin{eqnarray}
&&
\Delta a_{\mu} =  \frac{m_{\mu}^2}{ 4\pi^2 M_{\phi_2}^2 }\left[ -\frac{ M_{L}^3 }{m_{\mu} \Lambda^2}\left( \ln\left(\frac{M_L}{m_{\mu}}\right) -\frac{3}{4} \right) +\frac{1}{6} \right] \nonumber\\
\label{leptonmuon9}
\end{eqnarray}

Differently from the previous case, now we have a large $m_L$ enhancement. Currently limits range from 10GeV up to 100 GeV and largely depend on the search channel. For instance, L3 Collaboration has placed a limit of $M_L > 100$~GeV on a forth generation of charged leptons \cite{PDG}. It is not clear whether heavy charged leptons are attainable at the LHC, since Ref.\cite{DePree:2008st} states that those searches suffer from  large backgrounds, making difficult to pick a signal, whereas in Fig.6 of Ref.\cite{Ma:2014zda} we easily find $3\sigma$ and $5\sigma$ significance for $M_L=200-800$~GeV. In Ref.\cite{Kumar:2015tna} they claim one might possibly exclude masses up to 250GeV at the next LHC run if mixing between the heavy lepton and the SM tau is present. Regardless, ILC should definitely reach sensitive via the pair production of heavy leptons via Drell-Yann processes as discussed in Ref.\cite{Ari:2013wda,Altmannshofer:2013zba}.

Anyhow, the correction to g-2 turns out to be sizeable as we can see in Fig.\ref{plotinde}. Notice this is second most relevant contribution to g-2. Because of the negative sign, we can place a current and projected $1\sigma$ limit since the anomaly should be otherwise resolved. Taking $m_L=100$~GeV and $\Lambda = 1$~TeV, we derive $m_{\phi_2} > 2.8$TeV and $m_{\phi_2} > 4.8$TeV, using the current and projected sensitivity as shown in Fig.2. In Fig.3 we present the results for $\Lambda = 10$~TeV and $m_L=100$~GeV. The overall contribution is small, because of the large suppression imposed by $\Lambda$. Thus we impose $m_{\phi_2} > 150$GeV, using projected sensitivity.

\item Charged Scalar:

Charged scalars are evoked in several CHM and Technicolor models through the simplified lagrangian,

\begin{equation}
\mathcal{L}  \supset \lambda \frac{m_{\mu}}{\Lambda} \bar{\nu}_L \mu_R\, \phi^+ .
\end{equation}

The correction to g-2 appears in diagrams such as in Fig.1c, which results into,

\begin{eqnarray}
&&
\Delta a_{\mu}= \frac{1}{8\pi^2}\frac{m_\mu^2}{ M_{\phi^+}^2 } \int_0^1 dx \frac{ (m_{\mu} / \Lambda)^2( F_1(x) +  \ F_2 (x)) }{\epsilon^2 \lambda^2 (1-x)(1-\epsilon^{-2} x) + x}\nonumber\\
\label{scalarmuon4}
\end{eqnarray}where 
\begin{eqnarray}
F_1(x) & = &  -x (1-x)(x+\epsilon) \nonumber\\
F_2(x) & = & -x (1-x)(x-\epsilon)
\label{scalarmuon5}
\end{eqnarray}with $\epsilon = m_{\nu}/m_{\mu}$ and $\lambda= m_{\mu}/M_{\phi^+}$, which results in,
\begin{eqnarray}
&&
\Delta a_{\mu} = \frac{-m_\mu^2}{24\pi^2 M_{\phi^+}^2 }\frac{m_{\mu}^2}{\Lambda^2}
\label{scalarmuon6}
\end{eqnarray}

Notice that the overall correction is negative and quite dwindled due to the $m_{\mu}^4$ suppression as one can see in Figs.2-3, where we plotted the results for $\Lambda = 1$~TeV and $10$~TeV. We point out that there are various collider limits on mass of such singly charged scalars that lie in the $\sim 100-200$~GeV mass range \cite{chargedscalarlimits}. In specified UV models, stronger constraints might apply \cite{Freitas:2014pua}.

\item Charged Vector:

Sequential $W^{\prime}$ gauge bosons corrects the muon magnetic moment via the diagram in Fig.1d and lagrangian,

\begin{equation}
\mathcal{L}  \supset \frac{g}{\sqrt{2}} \bar{\nu} \mu_L\, W^{'+},
\end{equation}which results into,

\begin{eqnarray}
&&
\Delta a_{\mu} = \frac{m_\mu^2}{8\pi^2 M_{V^+}^2 } \left(\frac{g}{2\sqrt{2}}\right)^2 \int_0^1 dx \frac{ F_{1}(x) + \ F_{2} (x) }{\epsilon^2 \lambda^2 (1-x)(1-\epsilon^{-2} x) + x},\nonumber\\
\label{vectormuon4}
\end{eqnarray}where 
\begin{eqnarray}
F_1(x) & = &  2x^2(1+x-2\epsilon)+\lambda^2(1-\epsilon)^2 x(1-x)(x+\epsilon) \nonumber\\
F_2(x) & = &  2x^2(1+x+2\epsilon)+\lambda^2(1+\epsilon)^2 x(1-x)(x-\epsilon)\nonumber\\
\label{vectormuon5}
\end{eqnarray}with $\epsilon = m_{\nu}/m_{\mu}$ and $\lambda= m_{\mu}/M_{W^{\prime}}$.

\begin{eqnarray}
&&
\Delta a_{\mu} = \frac{10}{24\pi^2}\frac{m_\mu^2}{ M_{W^{\prime}}^2 } \left(\frac{g}{2\sqrt{2}}\right)^2
\label{vectormuon6}
\end{eqnarray}

One can clearly see that a singly charged vector boson rises as a natural candidate to explain the $(g-2)_{\mu}$ anomaly because it gives always positive contributions and for couplings of order one as we expect from gauge couplings, singly charged vector with masses of $\sim 400$~GeV might account for the anomaly as exhibited in Figs.2-3.  

However, searches in the regime where this new charged boson interacts only with right handed neutrinos, i.e when $g_{a10}=-g_{v10}$ give a 95\% C.L bound from LEP using effective operators which reads $g_{v10}/M_{W^{\prime}} < 4.8 \times 10^{-3} {\rm GeV^{-1}}$ \cite{Freitas:2014pua}, not still rulling out the region of parameter space which a $W^{\prime}$ accommodates g-2. Although, LHC data we can exclude $M_W^{\prime} > 2.55$~TeV at 95\% C.L, assuming SM coupling with fermions \cite{Aad:2011yg,Aad:2012dm}. The latter, literally rules out sequential singly charged gauge bosons as an alternative to address g-2.

\item Doubly Charged Scalar:

Doubly-charged scalars are typically present in models with triplet of scalars such as 3-3-1 models. There are two diagrams that give rise to correction to the muon magnetic moment: one when a photon is emitted from the doubly charged and another when the photon stems from the muon, or an exotic fermion. The lagrangian representing this contribution is,

\begin{equation}
\mathcal{L}  \supset  \frac{m_L}{\Lambda} \bar{L^c}\mu_R \phi^{++}
\end{equation}
\begin{eqnarray}
\Delta a_{\mu} (\phi^{++})& = & \frac{-q_H}{4 \pi^2} \left( \frac{m_\mu}{M_{\phi^{++}}}\right)^2 \left(\frac{m_L}{\Lambda} \right)^2 \times \nonumber\\
&& \int^1_0 dx \frac{ 2(x^3-x^2)}{ \lambda_1^2 x^2 + (1-\lambda_1^2)x + \lambda_2^2 (1-x) }+\nonumber\\
& &  \frac{-q_f}{8 \pi^2} \left( \frac{ m_\mu}{M_{\phi^{++}}}\right)^2 \left(\frac{m_L}{\Lambda} \right)^2 \times \nonumber\\
&& \int^1_0 dx \frac{ 2(x^2-x^3) }{ \lambda_1^2 x^2 +(\lambda_2^2-\lambda_1^2) x+ (1-x)}
\label{scalarmuon7}
\end{eqnarray}where $\lambda_1 = m_{\mu}/M_{\phi^{++}}$,$\lambda_2 = m_{L}/M_{\phi^{++}}$, $q_H=-2$ is the electric charge of the doubly charged scalar running in the loop, and $q_f=1$ is the electric charge of the muon in the loop. In the regime $M_{\phi^{++}} \gg m_{\mu},m_L$, this integral expression simplifies to,
\begin{equation}
\Delta a_{\mu}(\phi^{++})= \frac{-2}{3\pi^2} \left( \frac{m_L }{\Lambda} \right)^2 \left( \frac{m_{\mu} }{ M_{\phi^{++}} } \right)^2
\label{scalarmuon9}
\end{equation}
In Figs.2-3 we plotted the results for $m_L=100$~GeV and $\Lambda=1$ and $10$~TeV. It is clear the correction from a doubly charged scalar is negative and dwindled. See Refs. \cite{dchargedscalarlimits} for collider bounds on doubly charged scalars. 

\item Doubly Charged Vector:

The presence of doubly charged vectors a distinct feature of the so called 3-3-1 models, which might also have dynamic symmetry breaking in the context of Technicolor such as in Ref.\cite{Doff:2010pp,Doff:2007yu}. Massive gauge bosons in general have both vector and axial coupling. However, the vector component of the charged current involving two identical fields is null. In the model we will discuss further, the doubly charged gauge boson couples to the muon via an exotic heavy lepton as ,

\begin{equation}
\mathcal{L}  \supset \frac{g}{\sqrt{2}} \bar{L^c}\gamma^{\mu} \mu_L U^{++}
\end{equation}

In this case the integral is more complicated because the charged lepton mass can be comparable to the doubly charged boson one, plus the vector current is no longer null. One needs to solve the master integral below numerically for find the precise correction to g-2. Although, when the doubly charged gauge boson is much heavier than the muon and the exotic lepton we find,
\begin{eqnarray}
& &\Delta a_{\mu} =\left(\frac{g}{2\sqrt{2}}\right)^2\left[ \frac{m_\mu^2}{\pi^2 M_{V^+}^2 } \int_0^1 dx \frac{ \ F_{1}(x) +\ F_{2} (x) }{\epsilon^2 \lambda^2 (1-x)(1-\epsilon^{-2} x) + x} \right. \nonumber\\
&  &\left. -\left(\frac{g}{2\sqrt{2}}\right)^2\int_0^1 dx \frac{ F_{3}(x)+ F_{4}(x) }{(1-x)(1-\lambda^2 x) + \epsilon^2 \lambda^2 x} \right],
\label{Vcontri}
\end{eqnarray}where $\epsilon = m_{L}/m_{\mu}$ and $\lambda = m_{\mu}/M_{U^{\pm \pm}}$, and
\begin{eqnarray}
F_{1}(x) & = &  2x^2(1+x-2\epsilon)+\lambda^2(1-\epsilon)^2 x(1-x)(x+\epsilon) \nonumber\\
F_{2}(x) & = &  2x^2(1+x+2\epsilon)+\lambda^2(1+\epsilon)^2 x(1-x)(x-\epsilon),\nonumber\\
F_{3}(x) & = &  2x (1-x)(x-2(1-\epsilon))+\lambda^2 x^2 (1-\epsilon)^2(1+\epsilon-x) \nonumber\\
F_{4}(x) & = &  2x (1-x)(x-2(1+\epsilon))+\lambda^2 x^2 (1+\epsilon)^2(1-\epsilon-x).\nonumber\\
\label{vectormuon5}
\end{eqnarray}
\end{itemize}

Solving Eq.\ref{Vcontri} for $m_L=100$~GeV we find the numerical result in Figs.2-3. The result is insensitive to the scale of symmetry breaking differently from the previous cases since this is gauge interaction. It is visible from Figs.2-3 that the doubly charged boson induces the largest corrections to the muon magnetic moment. 

In summary, we have presented several simplified lagrangians applicable to several CHM and TCM. Now we apply our results to a technicolor model that extends the electroweak sector of the standard model.

\begin{figure}[!h]
\centering
\includegraphics[scale=0.5]{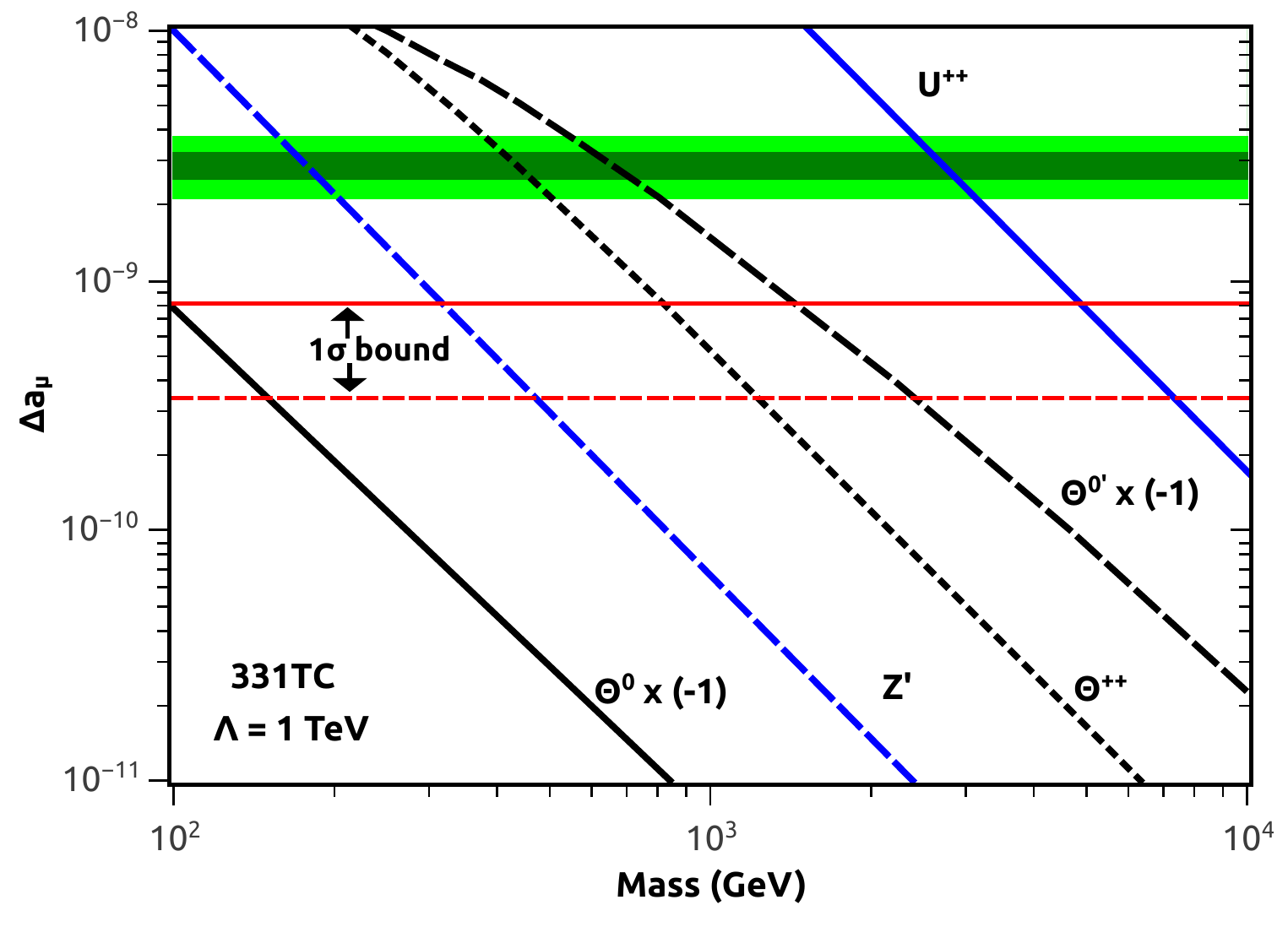}
\caption{Individual corrections to the muon magnetic moment as function of the boson masses for $\Lambda = 1$~TeV. The green band delimits the current and projected sensitivity of the muon magnetic moment. See text for detail.}
\end{figure}
\begin{figure}[!h]
\centering
\subfigure[\label{feymann2}]{\includegraphics[scale=0.5]{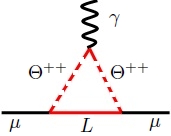}}
\subfigure[\label{feymann3}]{\includegraphics[scale=0.5]{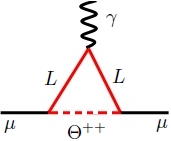}}
\subfigure[\label{feymann6}]{\includegraphics[scale=0.5]{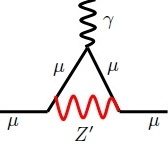}}
\caption{Feynmann diagrams arising in 331 models studied here.}
\end{figure}
\section{A Technicolor Model }
The model we briefly discuss below has been proposed in Ref.\cite{Doff:2012uq,Doff:2013wba} and is based on the gauge symmetry $SU(3)\otimes SU(3)_L\otimes U(1)_N$ that has been extensively studied in the literature \cite{Queiroz:2010rj,Mizukoshi:2010ky,Alves:2011kc,
Cogollo:2012ek,Alvares:2012qv,Alves:2012yp,
Caetano:2013nya}. The Technicolor model we investigate in this work is inspired by the known 3-3-1 minimal model, and therefore it inherits several features of the latter including the absence of dark matter particles. Nevertheless, dark matter can be incorporated with singlet fermions with no prejudice to our reasoning \cite{deS.Pires:2010fu} in agreement with recent measurements from WMAP9 and PLANCK\cite{Hooper:2011aj}. Anyways, in order to make the model anomaly free two of the three quark generations transform as ${\bf 3^*}$,  the third quark  family and the three leptons generations  transform as ${\bf 3}$. In the TC sector the triangular anomaly cancels between the two generations of technifermions, where technifermions  are singlets of $SU(3)_c$.

 The 331-TC model considered in this section presents the formation of two scales namely, the 331 symmetry breaking scale, $F_{\Pi} \sim  TeV$, and the TC scale $F_{TC} \sim 250GeV$ and the 331-TC model corresponds to an example of  two-scale Technicolor (TC) model.  The 331 symmetry breaking is implemented by the $U(1)_X$ condensate $\langle \bar{T}T \rangle$, that defines the mass scale of the exotic bosons ($Z^{\prime}$, $U^{\pm\pm}$) and the TC sector is responsible for the electroweak symmetry breaking. The contribution of the condensate $\langle \bar{T}T \rangle$ , mediated  by  Extended Technicolor interactions(ETC),  to  exotic pseudo goldstones \cite{Doff:2010pp} ($\Theta^{\pm\pm},\Theta^{'0}$)  masses can be estimated as 
\be 
M_{\Theta} \sim \frac{\langle \bar{T}T \rangle}{\Lambda^2_{ETC}(3)} \sim  {\rm few} (TeV).
\ee
\noindent  As a result these bosons in principle can acquire masses that are large enough to have escaped detection at the present accelerator energies. The contribution of $\langle \bar{T}T \rangle$ to mass of pseudo-scalar mimics the contribution expected by walking TC dynamics and the contribution of the TC sector for S parameter can be estimated as $S \sim 0.1$. Similarly to the Farhi-Susskind model \cite{Farhi}, the couplings of the neutral PGBs with muons are found to be \cite{Yue:2002pk},
\br 
{\cal{L}} = && i\frac{m_{L}}{F_{\Theta}\sqrt{2}}\left[ \bar{\mu}_{L} \Theta^{--} L^+_R + \bar{L^+}_L \Theta^{++}\mu_{R} + \bar{\mu}_L \gamma_5 \Theta^{'0} L_R \right]  \nonumber \\
\,\,\,\,+ && i\frac{m_{\mu}}{F_{TC}\sqrt{2}}\left[\bar{\mu}\gamma_5 \mu \Theta^{0} + \bar{\mu}_{L}\Theta^{-} \nu_{\mu R} + \bar{\nu}_{\mu L}\Theta^{+}\mu_{R}\right]
\er
\noindent where $F_{\Theta}=F_{\Pi} \sim  1 TeV $ is the decay constant of 331-TC (PGBs), $m_{L}$ corresponds to mass of exotic leptons. Combining Eq.26 with the doubly charged gauge boson correction shown in Eq.21, and the $Z^{\prime}$ correction derived in Eq.32, we have the total correction to the muon magnetic moment rising from the 331-TC model. The $Z^{\prime}$ contribution has been obtained from the neutral current \cite{Montero:1992jk},

\br
{\cal{L}} =\bar{\mu}\gamma^{\mu}(V_{\mu} - A_{\mu}\gamma_5)\mu Z^{'\mu}
\er
\noindent where 
\br 
&& V = -\frac{g}{4 cos(\theta_{W})}(L_{\mu} + R_{\mu}) \nonumber \\
&& A =  \frac{g}{4 cos(\theta_{W})}(L_{\mu} - R_{\mu})
\er 
\noindent  with
\br 
&& L = \frac{1}{\sqrt{3}}(1 - 4\sin^2\theta_{W})^{\frac{1}{2}} \nonumber \\
&& R = -2\sin^2\theta_{W}\left(\frac{3}{1 - 4\sin^2\theta_{W}}\right)^{\frac{1}{2}}. 
\er 

Neutral gauge bosons contribute to g-2 through the general master integral, 
leading to,

\begin{eqnarray}
&&
\Delta a_{\mu} (Z^{\prime}) = \frac{m_\mu^2}{8\pi^2 M_Z^{\prime 2} }\int_0^1 dx \frac{V^2 F_1(x)+ A^2 F_{2}(x) }{(1-x)(1-\lambda^2 x) +\lambda^2 x},
\label{vectormuon1}
\end{eqnarray}where, V and A and the respective vector and axial couplings and,

\begin{eqnarray}
F_1(x) & = & 2 x^2 (1-x) \nonumber\\
F_2(x) & = & 2 x(1-x)\cdot (x-4)- 4\lambda^2 \cdot x^3.
\label{vectormuon2}
\end{eqnarray}
In the limit $M_{Z^{\prime}} \gg m_{\mu}$ the integral simplifies to, 
\begin{equation}
\Delta a_{\mu}(Z^{\prime}) = \frac{m_{\mu}^2}{4 \pi^2 M_Z^{\prime 2}}\left(\frac{1}{3}V^2 - \frac{5}{3}A^2\right)
\label{vectormuon3}
\end{equation}with V and A given in Eq.31.
In summary to model corrects the muon anomalous magnetic through:\\
(i) $Z^{\prime}$\, (Eq.31)\\
(ii) $U^{++}$\, (Eq.21)\\
(ii) $\Theta^0$\, (Eq.25) \\
(iii)$ \Theta^{0\prime}$\, (Eq.25)\\
(iv) $\Theta^{++}$\, (Eq.25)\\

In Fig.4 we exhibit the individual contributions to g-2. Doubly charged gauge bosons give rise to the largest contribution, yielding the same constraints discussed in the previous section. Additionally, from Fig.4 we see that $\Theta^{0\prime}$ results into a sizeable and negative correction to g-2, whereas the doubly charged scalar induces a less relevant but positive one. Doubly charged scalar or singly charged scalars contributions are in general negligible. In this model it plays a more relevant role simply because of the $m_L$ enhancement, which is absent other 3-3-1 models \cite{Kelso:2013zfa,Kelso:2014qka,Profumo:2013sca,Kelso:2013nwa,Queiroz:2013lca,Cogollo:2013mga,Cogollo:2014jia,
Dong:2014wsa}. Since the doubly charged gauge boson is overwhelmingly more relevant than the others we can conclude that for $M_U^{++} \sim 2-3$~TeV the 3-3-1 TC can accommodate the g-2 anomaly with no prejudice to current bounds, once its contribution does not depend on the scale of symmetry breaking. In other words, we can push the scale of symmetry breaking to sufficiently high energies to obey electroweak and collider limits.

\section{Conclusions}

We have derived news physics contributions to the muon anomalous magnetic moment motivated by Composite Higgs models and Technicolor and shown general analytical expressions to account for new corrections stemming from neutral gauge bosons, charged gauge bosons, neutral scalar, singly charged scalars, doubly charged scalars and exotic charged leptons. We outlined which particles are able to reproduce the excess as well as derived $1\sigma$ bounds in case the anomaly is otherwise resolved.  Moreover, we commented on electroweak and collider bounds. Lastly, for concreteness we apply our results to a particular Technicolor model which might accommodate the g-2 anomaly with TeV scale gauge boson masses.


\begin{acknowledgments}
The authors thank Yotam Soreq, James Barnard, Daniele Barducci for valuable discussions. AD and CS are partly supported by Conselho Nacional de Desenvolvimento Cient\'{\i}fico e Tecnol\'ogico (CNPq). 
\end{acknowledgments}


\end{document}